\documentclass[fleqn,usenatbib]{mnras}
\usepackage{newtxtext,newtxmath}
\usepackage{xcolor}

\usepackage[T1]{fontenc}
\usepackage[utf8]{inputenc}
\usepackage{ae,aecompl}
\usepackage{lipsum}

\usepackage{graphicx}	
\usepackage{amsmath}	
\usepackage{amssymb}	
\usepackage{url}
\usepackage[colorinlistoftodos,prependcaption,textsize=tiny]{todonotes}
\usepackage[font=footnotesize]{subfig}
\usepackage{soul}

\usepackage{tikz}
\usetikzlibrary{automata, positioning}
 
\title[Modelling IFU data with INLA]{Spatial field reconstruction with INLA: Application to  IFU galaxy data}

\author[COIN collaboration]{%
S. Gonz\'alez-Gait\'an$^{1}$\thanks{E-mail: gongsale@gmail.com},
R. S. de Souza$^{2}$,
A. Krone-Martins$^{3}$,
E. Cameron$^{4}$,
\newauthor
P. Coelho$^{5}$,
L. Galbany$^{6}$,
E. E. O. Ishida$^{7}$,
for the COIN collaboration 
\\
$^{1}$CENTRA/COSTAR, Instituto Superior T\'ecnico, Universidade de Lisboa, Av. Rovisco Pais 1, 1049-001 Lisboa, Portugal\\
$^{2}$Department of Physics \& Astronomy, University of North Carolina at Chapel Hill, Chapel Hill, NC 27599-3255, USA\\
$^{3}$CENTRA/SIM, Faculdade de Ci\^encias, Universidade de Lisboa, Ed. C8, Campo Grande, 1749-016, Lisboa, Portugal\\
$^{4}$Big Data Institute, Li Ka Shing Centre for Health Information and Discovery, University of Oxford, Old Road Campus,\\ Oxford, OX3 7LS, United Kingdom\\
$^{5}$Instituto de Astronomia, Geof\'isica e Ci\^encias Atmosf\'ericas, Universidade de S\~ao Paulo, S\~ao Paulo, SP, Brazil\\
$^{6}$PITT PACC, Department of Physics and Astronomy, University of Pittsburgh, Pittsburgh, PA 15260, USA \\
$^{7}$Universit\'e Clermont Auvergne, CNRS/IN2P3, LPC, F-63000 Clermont-Ferrand, France\\
}

\date{Accepted XXX. Received YYY; in original form ZZZ}

\pubyear{2018}

\begin{document}
\label{firstpage}
\pagerange{\pageref{firstpage}--\pageref{lastpage}}
\maketitle

\begin{abstract}
Astronomical observations of extended sources, such as cubes of integral field spectroscopy (IFS), encode auto-correlated spatial structures that cannot be optimally exploited by standard methodologies.  This work introduces a novel technique to model IFS datasets, which treats the observed galaxy properties as realizations of an unobserved Gaussian Markov random field. The method is computationally efficient, resilient to the presence of low-signal-to-noise regions, and uses an alternative to Markov Chain Monte Carlo for fast Bayesian inference --- the Integrated Nested Laplace Approximation (INLA). As a case study, we analyse 721 IFS data cubes of nearby galaxies from the CALIFA and PISCO surveys, for which we retrieve the maps of the following physical properties:  age, metallicity, mass and extinction. The proposed Bayesian approach, built on a generative representation of the galaxy properties, enables the creation of synthetic images, recovery of areas with bad pixels, and an increased power to detect structures in datasets subject to substantial noise and/or sparsity of sampling. A snippet code to reproduce the analysis of this paper is available in the \href{https://github.com/COINtoolbox/Galaxies\_INLA}{COIN toolbox}, together with the field reconstructions of the CALIFA and PISCO samples.
\end{abstract}




\section{Introduction}

Astrophysical processes span a rich hierarchy of physical scales through time and space, making the evolutionary histories of the smallest and largest structures closely connected. This interplay between local and global dynamics is manifest in the observed morphologies of extended astronomical sources, most notably galaxies \citep[e.g][]{deZeeuw02,Cappellari11,Bershady10,Blanc13,Brodie14}. These objects  have been studied in ever growing detail thanks to the advent of Integral-Field Units (IFUs).  IFUs combine simultaneous spectral and spatial information, enabling novel ways to probe the spatial structure of galaxy properties, such as chemical abundances and star formation histories.  Indeed, IFU observations have become an integral component of large surveys at both low and high redshifts (e.g. \citealt[SAURON,][]{Bacon01}; \citealt[SINS,][]{ForsterSchreiber06}; \citealt[SAMI,][]{Croom12}; \citealt[MANGA,][]{Bundy15}; \citealt[AMUSING,][]{Galbany16}). 

Stellar population properties are traditionally obtained from IFU cubes using spectrum fitting codes in which each spaxel\footnote{A spaxel is a spectral pixel, i.e. every spatial pixel has an associated spectrum. In this work we use the terms pixel and spaxel interchangeably.} is treated independently \citep[e.g.][]{pPXF,STARLIGHT,STECKMAP,walcher+06,ULYSS,Sanchez16,Leja17}.  The integrated spectrum is modelled as a combination of simple stellar population libraries \citep{BC03,coelho+07,conroy_gunn10,maraston_stromback11,vazdekis+15,walcher+09}, enabling to retrieve a set of astrophysical parameters, e.g. ages, chemical abundances, global metallicities and other element ratios, or stellar masses and reddening \citep[see][for a recent review]{walcher+11}.  

But by underestimating astrophysical and/or instrumental correlations\footnote{From an instrumental viewpoint, fibre-bundle IFUs normally have gaps between fibres, so that usually 2-3 dithered observations are carried out to cover the gaps. Each reduced spaxel therefore contains data from different fibres, thus inducing an instrumental correlation among contiguous spaxels \citep{califa_dr2}.} between measured values at nearby locations, these approaches may fail to extract the complete signal from its noisy background.
This problem has been recognized before and several attempts to include spatial information have been made. These include global parametric models of the spatial distribution of physical quantities---adopted, for instance, in kinematic studies \citep[e.g.][]{Krajnovic06,watkins+13}---or the introduction of automatic spatial segmentations. The latter consists of using adaptive binning algorithms to form groups of contiguous spaxels over which measurements may be averaged to increase the signal-to-noise (S/N) ratio of the spectral information  \citep{Sanders01,Diehl06,cappellari_copin2003,Sanchez16}. The most common of these adaptive binning techniques is Voronoi tessellation \citep{Voronoi}, although variations based on intensity or isocontours have also been explored \citep{Sanders06,Sanchez12}. Most recently, \citet{batman} introduced a statistical segmentation technique that groups neighbouring regions by similarity across multiple properties.  Although segmentation approaches boost the statistical confidence of the spectral analysis, by averaging the signal over larger partitions, they do so at the expense of spatial precision \citep[e.g][]{cid+14}. This may lead to biased estimates if the averaged regions are inhomogeneous.

To mitigate the limitations of current segmentation-based approaches in astronomy, we introduce  a tool to take advantage of the spatial information encoded in IFU spectra. This algorithm profits from the presently available spectral synthesis and fitting codes and at the same time avoids the complexity and computational cost of performing a fully spatially resolved spectral synthesis and IFU fitting. The method relies on the use of Gaussian Markov Random Fields (GMRF) to estimate the unobserved spatial field. GMRF is a common approach for spatial modelling due to its simplicity, since it is completely determined through its mean and covariance structure. The conceptual motivation for the use of GMRF is to treat the data as a discrete and noisy sampling of an underlying continuous field, which the spatial model aims to reconstruct (see Section~\ref{sec:stats}).
For fast Bayesian inference, the model is implemented using the Integrated Nested Laplace Approximation \citep[INLA;][]{inlareview} framework embedded in the {\sc r-inla} package\footnote{\url{http://www.r-inla.org/}}. To make the approach seamless adaptable to traditional spectral synthesis codes, we apply it here to the existing maps of stellar population properties previously obtained through pixel-wise spectral fitting.

The structure of this paper is as follows. 
In section \ref{sec:data} we describe the IFU data from CALIFA and PISCO, and the respective stellar population fits with {\sc starlight}. Section \ref{sec:stats} introduces the underlying spatial statistical methodology, and in section \ref{sec:app} we discuss its application to the IFU data. Section \ref{sec:comp} compares our approach to the previous literature, section \ref{sec:applications} shows some applications, and we summarize our results in section \ref{sec:conc}.

\section{Optical IFU Data: CALIFA and PISCO}
\label{sec:data}
We retrieved data from the Calar Alto Legacy Integral Field Area survey \citep[CALIFA\footnote{\url{http://www.caha.es/CALIFA/public_html/}},][]{califa_survey,califa_dr1,califa_dr2,califa_dr3}, composed of data cubes and science data products of  667 nearby ($z < 0.03$) galaxies taken with the Potsdam Multi Aperture Spectrograph \citep[PMAS,][]{Roth05} in the PPak mode \citep{Verheijen04,Kelz06}. To this set, we add a sample of 104 galaxies from the PMAS/PPak Integral-field Supernova hosts COmpilation \citep[PISCO,][]{Galbany17} observed with the same instrumental configuration. All spectra cover the wavelength range 3745 -- 7500{\,\AA} with a resolution of R $\sim$ 850 (V500 mode) and a fraction with R $\sim$ 1650 (V1200 mode) in the blue (3400 -- 4840{\,\AA}).
 The CALIFA sample has been characterized in \citet{walcher+14}, and covers galaxies between $10^9$ and $10^{11.5}$\,M$_\odot$, with a peak between $10^{10}$ and 2 $\times$ $10^{11}$M$_\odot$ (see their Fig. 14). The morphologies are diverse, ranging from elliptical to late-type spirals, with a dominance of Sb and Sbc Hubble types. A large fraction of the galaxies (about 60\%; $\sim$450) show some emission lines, and \citet{walcher+14} estimated that the majority of those (43\%) are dominated by star formation (while 29\% are LINERs, 5\% are Seyferts, and the remaining lie in between the star formation and active galactic nucleus branches).

For the current analysis, we make use of the following stellar population parameters: age ($t [yr]$), metallicity ($Z$), mass ($M [M_\odot]$) and extinction in the visual band ($A_V$ [mag]).
Following \cite{2014A&A...572A..38G,2016A&A...591A..48G}, we estimated these parameters with the spectral synthesis package {\sc starlight} \citep{STARLIGHT,CidFernandes09} by fitting the spectrum at each spaxel with a linear combination of 248 simple stellar population (SSP) models from the ``Granada-Miles'' (GM) basis \citep{GonzalezDelgado15}. The GM consists of a regular grid of 62 ages spanning t = 0.001--14 Gyr and four metallicities ($Z/Z_{\odot}$ = 0.2, 0.4, 1, and 1.5), which are consistent with other commonly adopted bases such as \citet{Bruzual07}, as discussed  in \citet{cid+14}.
The final pixel-by-pixel age and metallicity are calculated from either the luminosity-weighted average of all SSP entering each fit: $\langle \log_{L}(t[\mathrm{yr}])\rangle$ and $\langle \log_{L}(Z)\rangle$, or the mass-weighted average: $\langle \log_{M}(t[\mathrm{yr}])\rangle$  and $\langle \log_{M}(Z)\rangle$. We only considered spectra for which the  S/N ratio in the continuum lay above 5 (estimated over the wavelength range of 4580-4640\AA). Additionally, we  also retrieved information from  the H$\alpha$ equivalent width (EW).

As a case in point, we illustrate the analysis  on the CALIFA galaxies, Scd NGC\,0309  and  E1 NGC\,0741, throughout the paper. The complete analysis of 721 galaxies observed with PMAS is presented in the \href{https://github.com/COINtoolbox/Galaxies\_INLA}{COIN toolbox}.
\subsection{Diagnostic  of IFU Spatial Correlation} 
\label{sec:Moran}

To illustrate the presence of spatial correlations in IFU cubes, this section provides an exploratory analysis using as an example the metallicity maps for the galaxies NGC\,0741 and NGC\,0309 from CALIFA. A simple approach to quantify the degree of spatial correlation is given by the  Moran autocorrelation coefficient \citep{Moran50}. This coefficient can be seen as an extension of the standard Pearson correlation \citep{Pearson01011895} for spatial data.

The general form of the Moran-I test statistic is:
\begin{equation}
I(h) = \frac{n}{\sum_{i=1}^{n}\sum_{j=1}^{n}w_{ij}}\frac{\sum_{i=1}^{n}\sum_{j=1}^{n}w_{ij}(x_i-\bar{x})(x_j-\bar{x})}{\sum_{i=1}^{n}(x_i-\bar{x})^2},
\end{equation}
where $n$ is the number of spatial units indexed by
$i$ and $j$, $\bar{x}$ is the mean value of the variable of interest, and  $w_{ij}$ is the weight  between observation pairs $i$ and $j$ separated by a certain  distance lag $h$. The correlation is taken between multiple pairs with similar separation, thus defining a global, and not a local, measure of the spatial correlation.

\begin{figure}
\includegraphics[width=0.95\columnwidth]{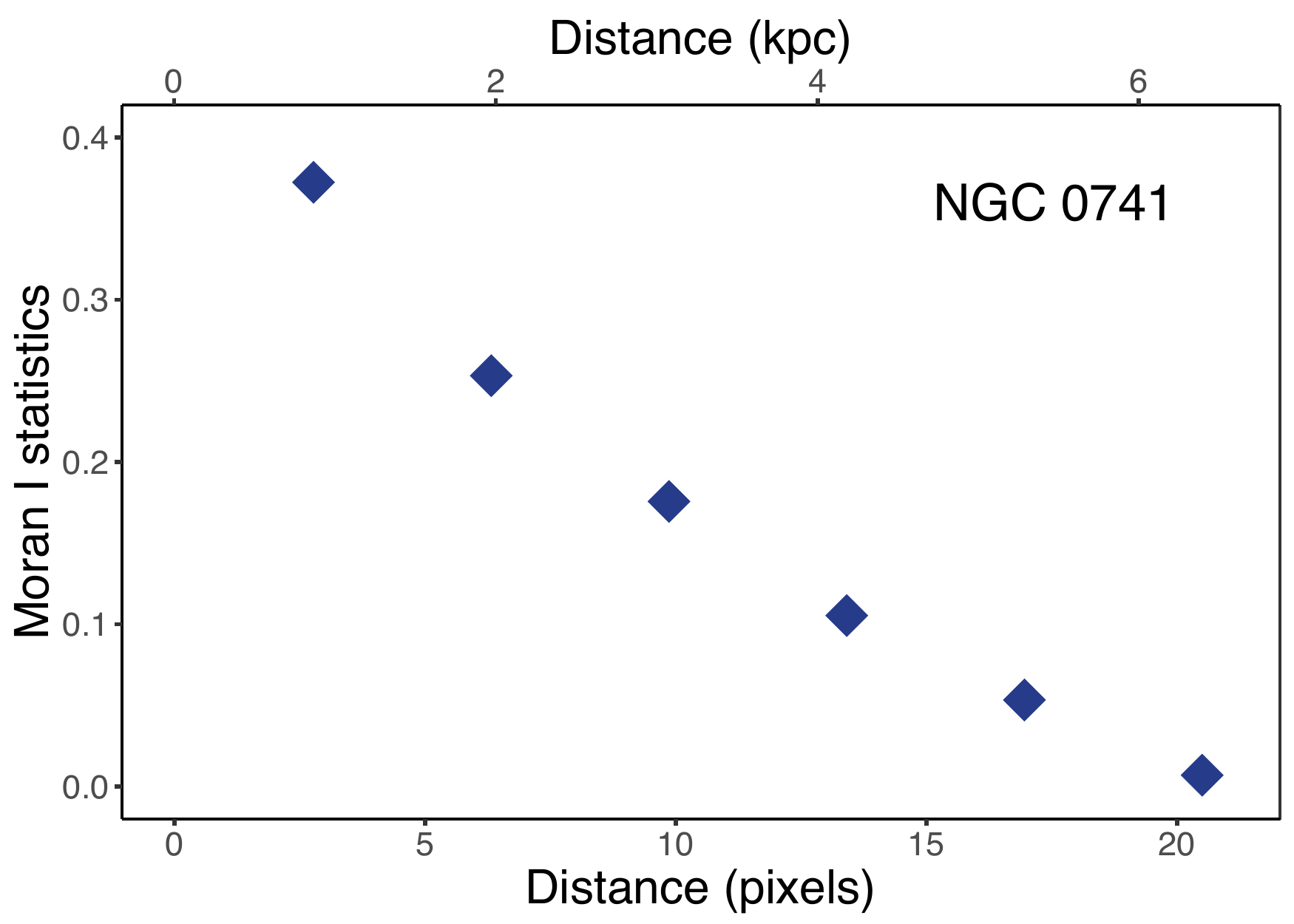}
\includegraphics[width=0.95\columnwidth]{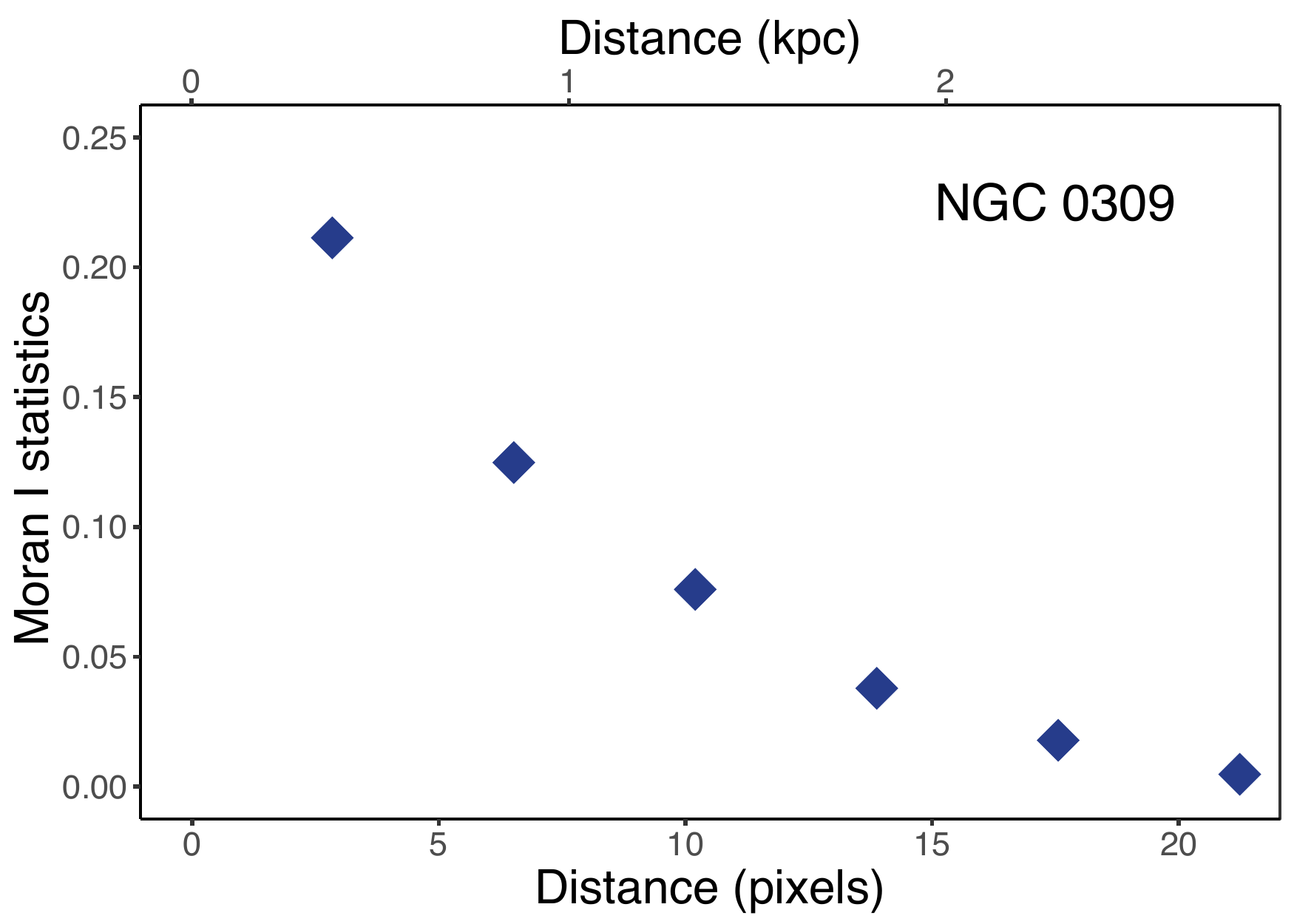}
\caption{Moran-I statistics as a function of pixel separation for metallicity maps of  NGC\,0741 and NGC\,0309. The approximate physical distance  displayed in the upper horizontal axis.}
\label{fig:moran}
\end{figure}

Figure \ref{fig:moran} shows the Moran-I test statistic for the metallicity maps of NGC\,0741 and NGC\,0309 as a function of pixel distances. The upper horizontal axis depicts approximate projected physical distance without taking into account galaxy inclination. The Moran-I statistic suggests that the level of correlation, as expected, is larger for nearby pixels and decreases as one moves away. The origin of this correlation can be either instrumental or physical but most likely a combination of both. But regardless of its origin, this simple example indicates that proper analysis of IFU cubes should account for the additional spatial dimension at all steps of the data analysis and of the physical properties inference process. By profiting from such additional information it should be possible to reconstruct the underlying scalar field of the data, and to infer if any fluctuations of this reconstruction are statistically significant.

\section{Spatial models via integrated nested Laplace approximation}
\label{sec:stats}

This section introduces few concepts of  Bayesian inference and spatial modelling to motivate its application to image/field reconstruction. For more details on Bayesian models, see e.g. \citet{2015bmps.book,blangiardo2015spatial,2017bmad,Sharma2017}.

Bayesian probability theory provides a mathematical framework to infer,  from the measured data,  the degree of plausibility of a given  model \citep[e.g.][]{jaynes2003probability}. It allows to update a current state of knowledge about a set of model parameters, $\theta$, in the face of new information. This is quantified by their posterior distribution, $p(\theta|y)$, i.e., the probability of the parameters, $\theta$, given the data, $y$. The Bayes's theorem lies in the foundation of the theory:
\begin{equation}
p(\theta|y) = \frac{\mathcal{L}(y|\theta)\pi(\theta)}{\int \mathcal{L}(y|\theta)\pi(\theta)d\theta}.
\label{eq:bayes2}
\end{equation}
The numerator on the right part of Equation~\ref{eq:bayes2} is composed by the model likelihood, $\mathcal{L}(y|\theta)$, i.e., the probability that the data, $y$, were obtained for given values of the model parameters, $\theta$, times the prior distribution, $\pi(\theta)$, which represents our state of knowledge before  accessing the new data. The normalization factor appearing in the denominator, called the evidence, describes the probability of obtaining the data considering all possible parameter values in the model.

For the task of image/field reconstruction, the problem is to ``infer" the underlying field, given an observed image/matrix. Since the ``image" in this case is a pixel matrix for which we have previously derived galaxy properties running a spectral synthesis analysis on each pixel, one may treat the entire matrix as a realization of an underlying random field of a given property. In other words, if treating an age map, the age at a given pixel is not treated as independent, but as spatially correlated to the estimated ages in nearby pixels. Therefore, the problem can be phrased from a spatial modelling perspective. 
To convey some intuition behind spatial models, let's start describing how to fit a Gaussian density distribution in a pixel-wise fashion, i.e. assuming spatial independence among pixels. For a given property $z$, normally distributed around the $i$-th pixel with mean value $\mu_i$, and variance $\sigma_{px}^2$  the model can be written as follows:
\begin{align}
\label{eq:spat1}
&z_i \sim \mathcal{N}(\eta_i,\sigma_{px}^2),\\
&\eta_i = \mu_i. \nonumber
\end{align}
Therefore, for each pixel, an independent  $\mu_i$ and $\sigma_{px}^2$ will be estimated, totally disregarding any information present in the surrounding pixels. The weakness of such an approach is evident since it will overreact to any noisy fluctuation present in the data, potentially leading to non-physical discontinuities in the map of galaxy properties.  Such effects are mitigated by segmentation based approaches \citep[e.g.][]{PyCASSO,batman}, but with the price of a greater resolution loss. 
To introduce a certain level of dependence among pixels, Equation (\ref{eq:spat1}) can be generalized as follows:
\begin{align}
&z_{ij} \sim \mathcal{N}(\eta_{ij},\sigma_{px}^2),\\
&\eta_{ij} = \mu_i + \kappa_{ij},\nonumber  
\end{align}
where $\kappa_{ij}$ encodes the spatial information between pixels $i$ and $j$. The strength and range of influence between the $i$-th and $j$-th pixels are usually  determined by the kernel function, and relies on expected properties of the underlying field (e.g. stationariness, isotropy).    
A common choice of  spatial kernel is the radial basis function (RBF) $\kappa_{ij} = \exp{(-\gamma||x_i-x_j||^2)}$, where $\gamma$ is a free hyper-parameter (commonly determined in the fitting process) and $x_i$, $x_j$ are pixel coordinates. The RBF kernel is stationary (translation invariant), isotropic (orientation invariant) and infinitely differentiable, thus inducing a higher degree of smoothness in the estimated field.

To account for the spatial autocorrelation of nearby galaxy regions, this work employs a type of spatially-discrete, continuously-valued distribution---the Gaussian Markov random field \citep[GMRF;][]{rue2005gaussian,Lindgren11}. GMRF are one of the simplest, and yet, efficient approximations of continuous random fields.  
A given random vector, $x$, is said to be distributed as a GMRF if it follows a multivariate normal distribution with Markov independence properties defined over a neighbourhood structure, $S_i$:
\begin{equation}
x_i \perp x_j\ |\ x_{k \in S_i}$ ($i\neq j$, $j \not\in S_i),
\end{equation}
where $\perp$ here denotes (conditional) independence, and the indexes represent spatial/pixel coordinates. 
The condition can be graphically represented as follows:
\begin{equation}
 \begin{tikzpicture}[font=\sffamily]
    \node[state,
          draw=none,
         fill=gray!40!white] (x) {$x_i$};
    \node[state,
          draw=none,
         fill=gray!30!white,
         right=2cm of x] (y) {$x_k$};
      \node[state,
          draw=none,
         fill=gray!10!white,
         right=2cm of y] (z) {$x_j$};        
    \draw[line width=0.5mm,
          auto=left]
        (x) edge[auto=left ] node {$S_i$} (y)
        (y) edge node {} (z);
   \end{tikzpicture}  
  \end{equation}
which is reflected in the relation $f(x_i,x_j,x_k)f(x_k) = f(x_i,x_k)f(x_j,x_k)$, where $f$ is a given probability density distribution. Intuitively it gives more strength to the neighbours, i.e. knowing $x_k$ renders $x_j$ less relevant for predicting $x_i$. This is also a type of partial pooling, i.e. the information is shared among nearby pixels to estimate the summary statistics within each pixel. 
The radius of influence of each pixel and its weight is encoded by the spatial kernel.

This neighbourhood structure implies a sparse precision (inverse covariance) matrix, $Q = \Sigma^{-1}$, for the corresponding density of $x \sim \mathcal{N}_\mathrm{GMRF}(\mu,\Sigma)$:
\begin{equation}
\mathcal{N}_\mathrm{GMRF}(\mu,Q) = (2\pi)^{-n/2}|Q|^{1/2}\exp{\left(-\frac{1}{2}(x-\mu)^{\prime}Q(x-\mu)\right)},
\end{equation}
\noindent where $\mu$ denotes the mean vector (usually set to zero).  The specific neighbourhood structure and precision matrix we use for the GMRF is that approximating a continuous Gaussian random field using a generalization of the RBF kernel, the  Mat\'ern covariance function \citep{Lindgren11}; namely, for a spatial separation, $h$, the covariance:
\begin{equation}
\Sigma(h) = \sigma^2\frac{2^{1-\nu}}{\Gamma(\nu)}(\kappa h)^\nu K_\nu(\kappa h),
\end{equation}
with $K_\nu(\cdot)$ denoting the modified Bessel function of second kind and order $\nu > 0$, and two parameters, $\sigma$ and $\kappa$, controlling the variance ($\sigma^2$) and the
range ($\approx \sqrt{8\nu}/\kappa$) at which the spatial correlation becomes negligible.

With the GMRF definition in place, our model for the spatial distribution of measured galaxy properties can written in Bayesian hierarchical form as follows.  For pixels, $i=1,\ldots,n$, at locations, $x_i$, with non-spatial estimates of e.g. metallicity (or age, mass, etc.), $z_i$, having quoted uncertainties, $\sigma_i$:
\begin{gather}
z_i \sim \mathcal{N}(f(x_i),\sigma_i^2),\nonumber\\
f(\cdot) = g(\cdot)+h(\cdot),\nonumber\\
g(\cdot) \sim \mathcal{N}_\mathrm{GMRF}[\sigma,\kappa],\\
h(\cdot) = \alpha + \beta\times\mathrm{edist}(\cdot),\nonumber\\
\sigma, \kappa, \alpha,\beta \sim \pi.\nonumber
\end{gather}\label{inlamodeleqn}
%
Here,  $f(\cdot)$ represents the latent (hidden/source) image composed as the sum of the GMRF over the 2D spatial plane and a linear function in the (mass-weighted, elliptical-aperture) radial distance (`edist'). Gaussian process models are rightly hailed for their impressive flexibility and can indeed reproduce a smooth radial trend given only a zero mean vector. In the case that one does have an expectation of any broad trends that might be represented parametrically, including these terms explicitly can improve predictive accuracy in the low signal-to-noise regime. Alternatively, it is possible to represent the radial component by an Ornstein-Uhlenbeck \citep[OU;][]{OU1930} process.\footnote{The OU process is the continuous analogue of the discrete autoregressive model \citep[see e.g.][chapter 10.10]{2017bmad}, and is a particular case of a Gaussian process with bounded variance and stationary probability distribution.}
This introduces a weaker tendency towards radial smoothing but with potentially even greater flexibility (e.g.\ for galaxies with a star forming core). We have also explored OU processes, but it does not lead any significant effects in our results.

The final layer of the hierarchical model denotes completion of the Bayesian framework with appropriate priors---given for now the notation for a place holder distribution, $\pi$---on the hyper-parameters of the GMRF and the parameters of the parametric component.  It is well known for Gaussian process regressions that the values these hyper-parameters can take have a strong influence on the smoothness of the resulting image estimates, yet they will often be poorly constrained by the likelihood function for the observed data \citep[e.g][]{Zhu06}. Hence, the importance of (i) careful prior choice and sensitivity analysis, and (ii) integration (or `marginalisation') over the posteriors for the hyper-parameters to propagate these uncertainties through to the predictive distribution.  An efficient computational implementation of this model structure and uncertainty analysis is facilitated by the {\sc r-inla} package.

{\sc r-inla}\footnote{\href{http://www.r-inla.org/}{http://www.r-inla.org/}} is a {\sc r} package designed for modelling spatial data. It has strong computational gains relative to a direct MCMC sampling strategy achieved over three fronts. Firstly, its use of an effective Laplace approximation \citep{tierney1986,Rue09} for estimation of the marginal likelihood of the high dimensional random field component, given the observed data, and a vector of trial hyper-parameters (see appendix \ref{app:Laplace} for a brief definition).  Secondly, the use of grid-based integration over the low dimensional space of hyper-parameters rather than Monte Carlo exploration \citep[][see also recent work in this direction on sequential quasi Monte Carlo; \citet{gerber2015}]{Rue09}.  Both of these steps trade a degree of bias for computational speed ups, their suitability being generally well regarded for the class of models  targeted by the INLA approach\footnote{Although this is not a universal consensus; see \citealt{taylor12} for a dissenting view with regard to modelling Poisson processes.}.  Lastly, 
he sparse precision structure of the GMRF component allows for much faster matrix operations than those for an equivalent dense matrix. 

The INLA methodology is very popular in a number of fields, and has been successfully applied to build spatio-temporal disease mapping models \citep{Schroedle11},  to modelling fish populations in the Northwest Atlantic Ocean \citep{Boudreau17}, to probe population size dynamics from molecular data \citep{Palacios12}, for mapping the global ozone levels \citep{Bolin11}, air pollution \citep{Cameletti13}, and forest fires in Portugal \citep{Natario14}, and predicting extreme rainfall events in space and time \citep{Opitz2018}, just to cite a few. 
Therefore, the case study herein employed -- spatial maps from IFU cubes -- represents just a glimpse of the potential of these spatial  models with INLA , and we advocated for their use in other astronomical problems. 

A final note on the modelling strategy adopted here. 
Due to the so-called `shrinkage' \citep{Copas83}\footnote{Intuitively, it assumes that a raw estimate (within pixel) can be improved by combining it with other information (inter-pixel).} effect,
the spatial representation in our approach allows for a sharing of information between nearby pixels. We expect this to yield more accurate predictions (in a mean squared error sense) than what is achievable through a pixel-by-pixel analysis.
This well-studied effect targeted by hierarchical Bayesian models \citep{efron1973} occurs naturally in high dimensional problems where a degree of correlation is anticipated \textit{a priori}, but (remarkably) can even  occur when no such correlation holds under the true underlying model \citep[][i.e., `Stein's phenomenon']{stein1956}.  As will be shown later, a fortunate side-effect is a more reader-friendly visual representation of the data.

Our full hierarchical  model structure composed of a GMRF, and optimized via INLA,   under the  {\sc r-inla} framework,  will henceforth be referred to simply as ``INLA'' for the sake of simplicity.

\section{Spatial analysis of IFU maps}
\label{sec:app}

This section  discusses the application of our statistical framework to the CALIFA galaxies, NGC 0309, a spiral galaxy, and NGC 0741, an elliptical galaxy. The first four raw (i.e. non-spatially modelled) quantities: luminosity-weighted age and metallicity, mass and extinction, were extracted individually and independently by {\sc starlight} runs on each spaxel of the IFU cubes, while the H$\alpha$ equivalent width measurements were obtained from independent Gaussian profile line fits. INLA was then used to estimate the underlying spatial distribution of the latent parameters as per the hierarchical model structure described above.

\begin{figure*}
\centering
  \begin{minipage}[l]{1.0\linewidth}
         \centering
         \includegraphics[trim={0 0.5cm 0 0.5cm },clip,width=0.41\columnwidth]{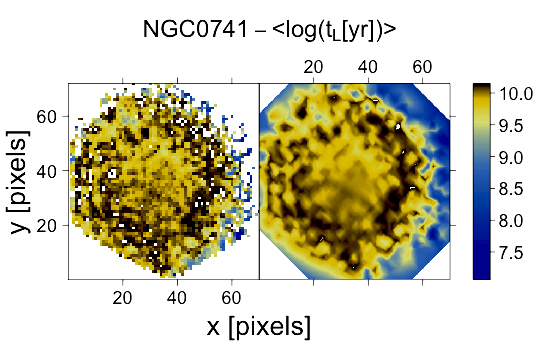}
         \includegraphics[trim={0 0.5cm 0 0.5cm },clip,width=0.41\columnwidth]{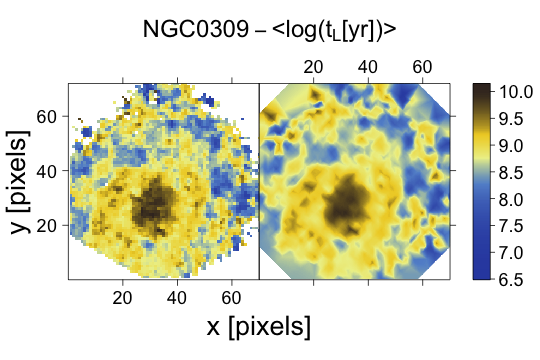}
  \end{minipage}

  \begin{minipage}[l]{1.0\linewidth}
         \centering
         \includegraphics[trim={0 0.5cm 0 0.5cm },clip,width=0.41\columnwidth]{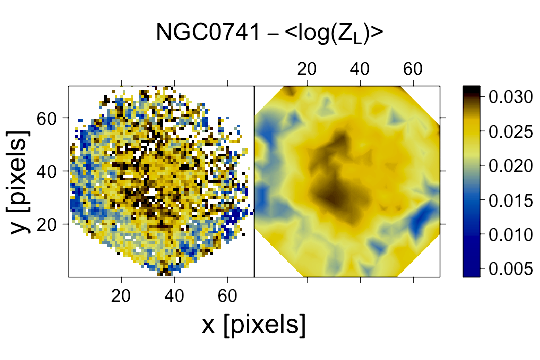}
         \includegraphics[trim={0 0.5cm 0 0.5cm },clip,width=0.41\columnwidth]{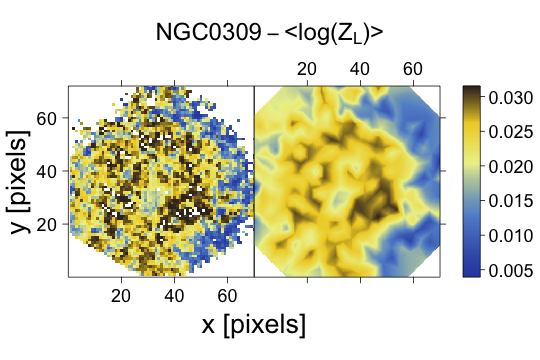}
  \end{minipage}
     
  \begin{minipage}[l]{1.0\linewidth}
         \centering
         \includegraphics[trim={0 0.5cm 0 0.5cm },clip,width=0.41\columnwidth]{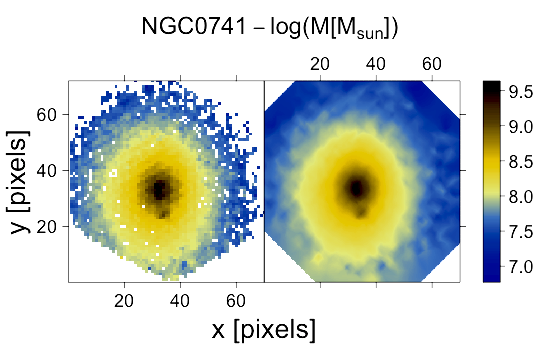}
         \includegraphics[trim={0 0.5cm 0 0.5cm },clip,width=0.41\columnwidth]{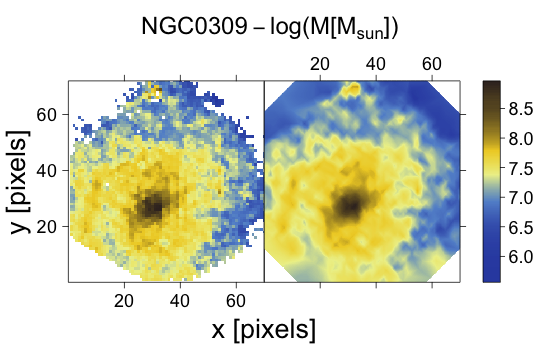}
   \end{minipage}
     
  \begin{minipage}[l]{1.0\linewidth}
         \centering
         \includegraphics[trim={0 0.5cm 0 0.5cm },clip,width=0.41\columnwidth]{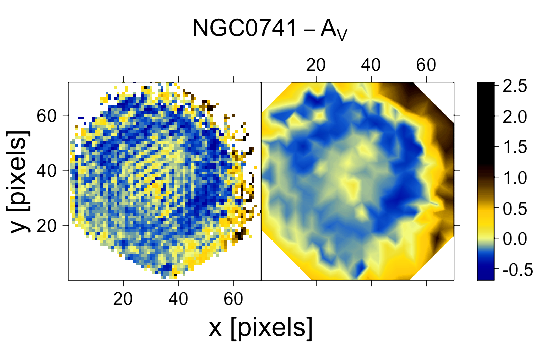}
         \includegraphics[trim={0 0.5cm 0 0.5cm },clip,width=0.41\columnwidth]{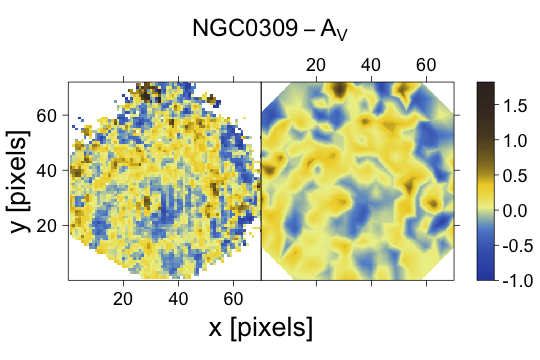}
  \end{minipage}
     
  \begin{minipage}[l]{1.0\linewidth}
         \centering
         \includegraphics[trim={0 0.5cm 0 0.5cm },clip,width=0.41\columnwidth]{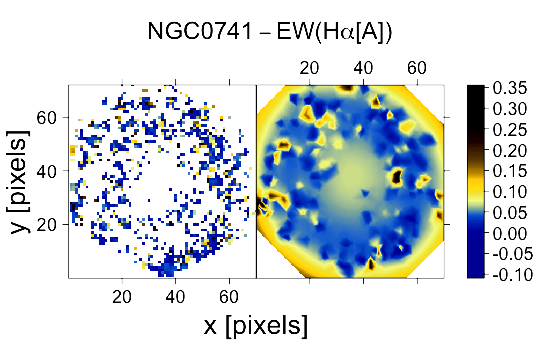}
         \includegraphics[trim={0 0.5cm 0 0.5cm },clip,width=0.41\columnwidth]{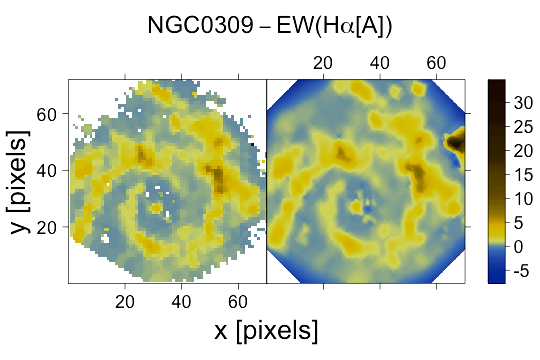}
   \end{minipage}

\caption{\label{inlaresults}From top to bottom: luminosity-weighted Age: $\langle\log(t_L[yr])\rangle$, Metallicity: $\langle\log(Z_L)\rangle$, Mass: $\langle\log(M[M_{\odot}])\rangle$, Reddening: $A_V$ and H$\alpha$ EW(\AA) maps obtained by directly plotting the results from the independent spaxel fits (left) and the INLA reconstructed parameter field (right). NGC\,0741 (CALIFA K0068) in the left-most figures. NGC\,0309 (CALIFA K0034) in the right-most figures.}
\end{figure*}

In Figure \ref{inlaresults} we present side-by-side comparisons of the original and resulting (INLA reconstructed) maps for each of those quantities. It is evident that both global and local structures in these galaxies are more easily seen in the reconstructed images than from the simple spatial plots of the (noisy) individually determined parameters. Furthermore, these results illustrate the predictive capability of the model to interpolate over empty spots in the galaxy original IFU data or in the outskirts of the instrument field-of-view, at positions where no individual {\sc starlight} fits were obtained due to excessively low signal-to-noise. This would also be the case if some region of the focal plane was missing coverage due to either instrumental defects (e.g. broken fibers) or to instrumental design (e.g. fiber gaps). We stress that this type of prediction is only possible because the physical information is not purely random: it is spatially correlated.

Furthermore, given the multiple dithering observations of the CALIFA cubes, if these are taken under different observing conditions, a known systematic spatial pattern can sometimes be present. The $A_V$ input maps show best this artificial structure which is then greatly reduced with INLA. Even though INLA is heavily dependent on the input data, large-scale systematics will be carried on to the INLA prediction but small enough systematics that are non-correlated will be safely ignored 

It is important to emphasize that the INLA prediction is not a smoothing convolution of the original data; if there are  spatial correlations in the data, it takes this information into account while predicting the underlying field. Nor does it degrade unnecessarily the image resolution by prior binning of the maps to increase the signal.

\begin{figure*}
	\begin{minipage}[l]{1.0\linewidth}
    \vspace*{0.4cm}
	\centering
		\includegraphics[trim={0 0 0 2cm},width=0.19\columnwidth]{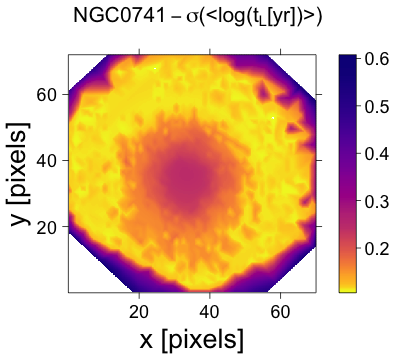}
		\includegraphics[trim={0 0 0 2cm},width=0.19\columnwidth]{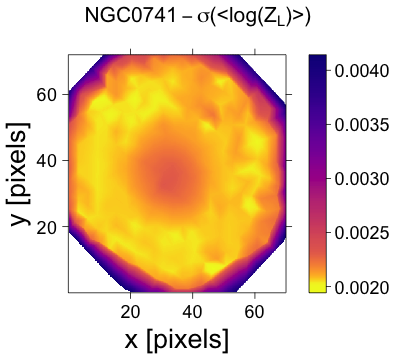}
		\includegraphics[trim={0 0 0 2cm},width=0.19\columnwidth]{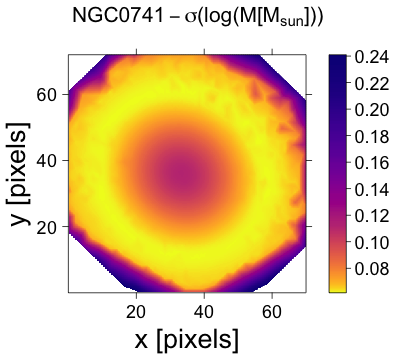}
		\includegraphics[trim={0 0 0 2cm},width=0.19\columnwidth]{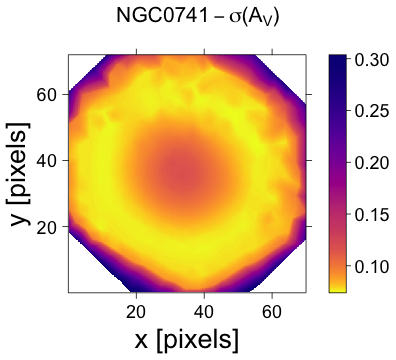}
		\includegraphics[trim={0 0 0 2cm},width=0.19\columnwidth]{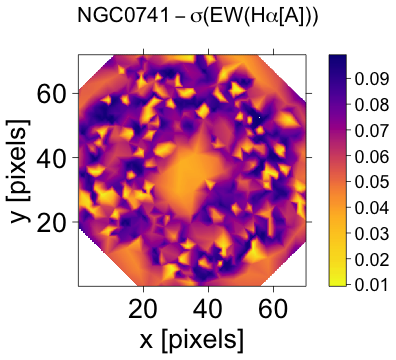}
	\end{minipage}
	\begin{minipage}[l]{1.0\linewidth}
	\centering    
		\includegraphics[trim={0 0 0 0cm},width=0.19\columnwidth]{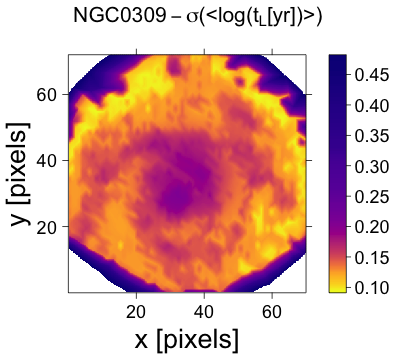}
   		\includegraphics[trim={0 0 0 0cm},width=0.19\columnwidth]{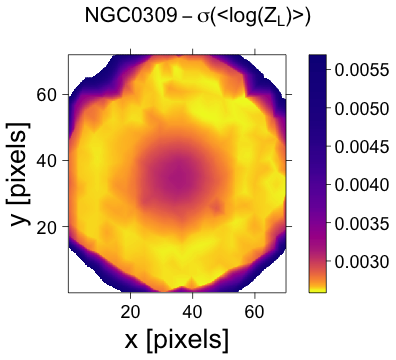}
		\includegraphics[trim={0 0 0 0cm},width=0.19\columnwidth]{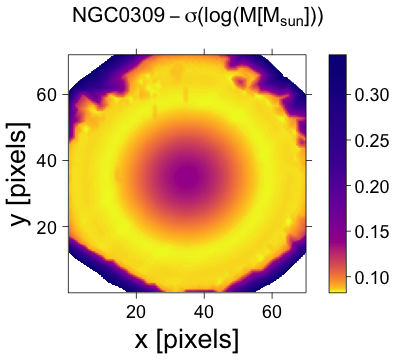}
		\includegraphics[trim={0 0 0 0cm},width=0.19\columnwidth]{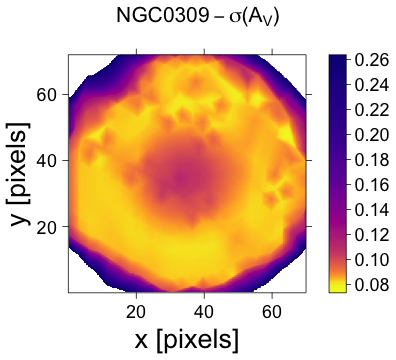}
		\includegraphics[trim={0 0 0 0cm},width=0.19\columnwidth]{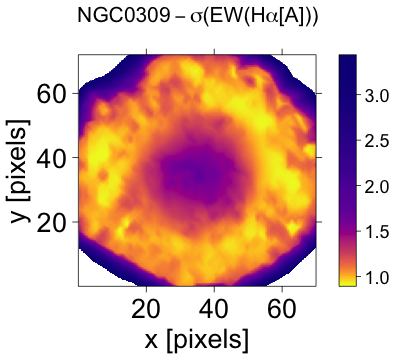}
	\end{minipage}
\caption{\label{inlaresultserrors} Map of the reconstruction uncertainty estimates derived by INLA for the  Age, Metallicity, Mass, Reddening and H$\alpha$ EW (from left to right) maps shown in Figure \ref{inlaresults}. NGC\,0741 (CALIFA K0068) in the top. NGC\,0309 (CALIFA K0034) in the bottom.}
\end{figure*}

Another considerable advantage of INLA is that it approximates the full posterior distribution --- accordingly, it is possible to assess the statistical significance of the results. The INLA maps shown in Figure \ref{inlaresults} represent the mean value of the posterior distributions at each pixel. Figure \ref{inlaresultserrors} shows the maps for the estimated standard deviation from the resulting posteriors. This information can be used as a first proxy for the prediction error, since the posterior distributions are approximately normal in the present case. For all properties of both galaxies, the predicted dispersion maps are consistent with expectations: larger uncertainties are obtained at lower signal-to-noise data. The regions of the input map that are most uncertain, usually due to lack of data (e.g. no {\sc starlight} fit was obtained) consistently have larger dispersions. This can be observed, for instance, in the outskirts of the maps, where the posteriors are being predicted beyond the original boundaries. It is also seen in the estimate for the H$\alpha$ EW(\AA) map of NGC 0741, that shows posteriors which are statistically compatible with null in a large fraction of the reconstructed scalar field. A possibly false emission is inferred in the centre of this galaxy, where there is no data. Nonetheless, the INLA error tells us that this is not significant and is consistent with zero. We checked that this feature disappears with a different input radial component ($h(\cdot)$ in eq.~\ref{inlamodeleqn}). We therefore caution that one should never interpret the results beyond what properly estimated errors allow.

A note on the computation of input uncertainties, the $\sigma_i$ of our hierarchical model (equation \ref{inlamodeleqn}). While the {\sc starlight} code by itself does not estimate uncertainties for the fitted parameters, \citealt{cid+14}, with the help of extensive simulations, estimated the uncertainties for different signal-to-noise values. This is the approach we adopt here although we have also tested an alternative strategy of allowing the uncertainties to be learnt entirely within the {\sc r-inla} code. In both cases, we obtained similar results indicating that the addition of the {\sc starlight} estimated uncertainty maps following the aforementioned technique is not an essential component here for effective inference.

We also note that the predicted maps from INLA have in fact a higher correlation at neighbouring pixels than the input maps (calculated with Moran statistics - see Section \ref{sec:Moran}) showing that the underlying covariance structure is lessened by the random noise introduced when relying in the pixel-by-pixel estimate. Having said that, one should always be careful with the choice of kernel in order to avoid undesired levels of smoothness.

\section{Comparison to other techniques}
\label{sec:comp}
To contextualize our technique we show here a qualitative comparison to alternative mapping approaches used in the literature. The most adopted procedures to analyse IFU data  are based on segmentation of the space by aggregating neighbouring spaxels into larger regions to optimize the signal-to-noise \citep[e.g.][]{PyCASSO}. This happens \emph{prior} to some model fit (e.g. using {\sc starlight}) that is performed inside each segmented region. INLA, on the other hand, is drastically different: it builds a full spatial map of a given property generally \emph{after} an initial model fit has been performed. Accordingly, INLA does not require to first degrade the spatial resolution of the original data.

\begin{figure*}
\includegraphics[trim=0 0.1cm 0 0.1cm,clip,width=0.95\linewidth]{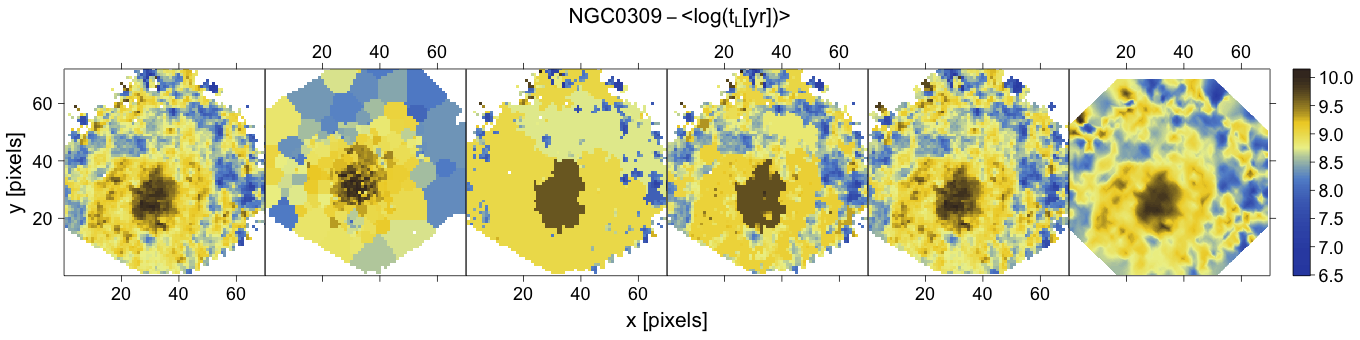}
\caption{\label{inlapycassostarlight} Comparison of  different spatial algorithms for NGC\,0309 (CALIFA K0034) age maps. From left to right: {\sc STARLIGHT} map obtained pixel-by-pixel, PyCASSO (i.e. {\sc starlight} on a \emph{previous} tessellated input image), {\sc BaTMAn} segmentation with two different priors affecting the tessellation regions (1781 and 3626 cells) \emph{after} the {\sc starlight} fit,  a  Gaussian smoother with a RBF kernel, and INLA model \emph{after} the age fit.}
\end{figure*}

Figure \ref{inlapycassostarlight} shows age maps of the galaxy NGC\,0309. The left-most panel is a representation of {\sc starlight} result fits, pixel-by-pixel. The second panel is the {\sc starlight} fit obtained on the raw data pre-processed with Voronoi tessellation to account for low S/N spaxels from PyCASSO\footnote{\url{http://www.pycasso.ufsc.br}} \citep{PyCASSO}. Voronoi tessellation is a division of the space into cells that contain all points equidistant to the nearest sites. The third and fourth panels are two results from {\sc BaTMAn} \citep{batman}. {\sc BaTMAn} is a more general Bayesian segmentation approach that partitions a plane based on similarities in one or more characteristics. Here we use it in its simplest form by finding similarities in the {\sc starlight} age maps. So, as opposed to the PyCASSO example, the segmentation is performed \emph{after} the fit. Finally, the last column shows INLA's prediction on the output age map of {\sc starlight}. %

The traditional approaches, such as the PyCASSO tessellation prior to the fit, erase a significant fraction of the spatial information and average the physical properties over large areas. INLA, on the other hand, does not a priori reduce the spatial resolution, since it constructs a model based on the data. In the case of the segmentation-based fit with {\sc BaTMAn}, the number of the resulting cells of the spatial map can vary depending on user-selected parameters (we show two examples in Figure~\ref{inlapycassostarlight}); but we note that it can never reach  a resolution as high as the original pixel-to-pixel {\sc starlight} fit; and it does not have the capacity of estimating values in regions without data.  An innovation of {\sc BaTMAn} is its capability to include more than one dimension (i.e. one parameter) in the partition of cells. Although not addressed in the present work, given the hierarchical structure of our model (equation \ref{inlamodeleqn}),  it  can be easily extended to include multiple  covariates,  and the correlations among them.

%
Figure \ref{inlapycassostarlight} clearly indicates qualitatively that traditional segmentation techniques are not capable of including spatial correlations consistently. They operate on a trade-off between a gain in S/N at a significant cost of lowering the spatial resolution of the data. 
In contrast, INLA generates a self-consistent spatial model that inter-relates neighbouring information by considering a spatial correlation matrix. In principle, one could envisage extreme scenarios where it could become advantageous to use both types of algorithms: if there are very large regions where no fit is feasible because of extremely low S/N, a preliminary segmentation could be beneficial for a first fit to constrain the prior distributions; these would then be used during the posterior reconstruction of the final fit with INLA. However, given the loss of spatial information in the binning algorithms, this type of procedure should only be recommended in the most extreme cases.
In addition, we also provide a comparison against a simple Gaussian smoother with  a RBF kernel using the  {\sc r} package {\sc fields} \citep{fields}. A visual inspection confirms that our approach  outperforms a naive interpolation.

\section{Further potential}
\label{sec:applications}

Although it is out of the scope of this work to provide an exhaustive list of the model capabilities, we discuss below some scenarios in which these techniques can be valuable.  This section provides three potential applications of our approach: i) Construction of probabilistic maps of galaxy properties; ii) Prediction of under-sampled or missing regions (inpainting); iii) Shared reconstruction of multiple fields.  

\subsection{Probabilistic maps}

\begin{figure}
\centering
\includegraphics[trim=0.1cm 0.1cm 0.1cm 0.1cm,clip,width=0.95\linewidth]{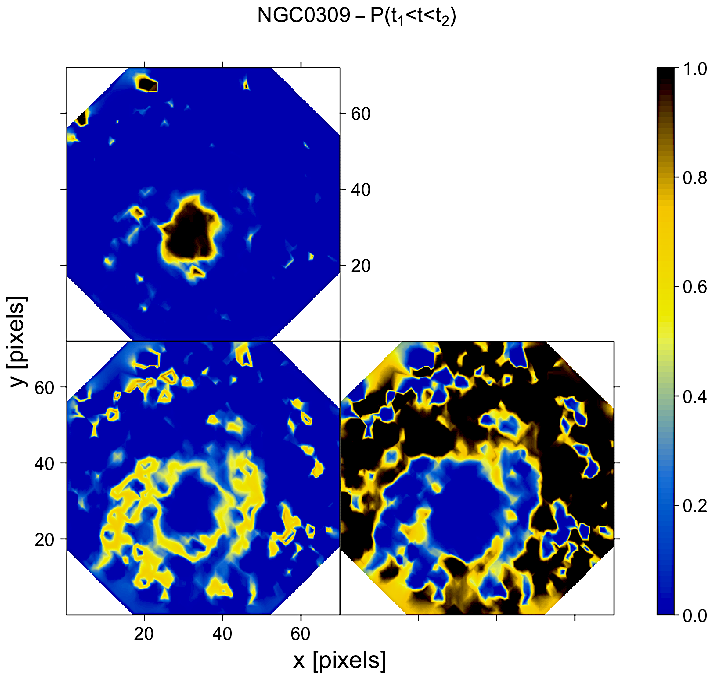}
\caption{Probability maps of NGC\,0309 for three ranges  of age, $\log(t[yr])$, arranged anticlockwise: $\log(t[yr]) > 9.5$, $9.3 > \log(t[yr]) > 9.0$ and $\log(t[yr] < 8.9$. The bins were chosen to represent  bottom ($ < 2.5\%$), middle ($ 32\% - 68\%$) and top ($> 97.5\%$) quantiles of the  reconstructed population age map (see Figure \ref{inlaresults}).}  
\label{fig:prob_age}
\end{figure}

The probabilistic nature of the Bayesian spatial model  enables the production of  credible maps of galaxy properties for a given range of astrophysical interest. This is simply done by integrating the  posterior over the interval of choice. 
Figure \ref{fig:prob_age} highlights three regimes of ages for  NGC\,0309: $\log(t[yr]) \geq 9.5$, $9.3 > \log(t[yr]) \geq 9.0$, and $\log(t[yr] < 8.9$, which are chosen to match regions between  3-quantiles  of the mean age map shown in figure  \ref{inlaresults}.

A visual inspection clearly discriminates the locus of  higher probability to find old stellar populations, concentrated around the bulge of the galaxy, in contrast to  younger populations, more likely to be found as one moves radially outwards from the centre. Although expected, the formalism allows to place probabilistic  intervals  on this physical statement.  For instance, that stellar populations within 500 pc ($\sim$ 5 pixels) from the centre are older than $10^{9.5}yr$ with 95\% probability. 

\subsection{Missing data}

The presence of missing data is ubiquitous in astronomical datasets, either due to data corruption or due to the impossibility of acquiring a large number of telescope-time expensive observations. 

One common scenario of data corruption is the presence of bad pixels in photon detectors. Other common situations are the need to interpolate over regions under the presence of contaminant signals: cosmic rays; foreground stars; the Milky Way foreground emission; artificial satellite and space-debris tracks; complex background patterns, etc. Inpainting methods are widely adopted to address these issues, from simple local interpolation \citep[e.g.][]{2001PASP..113.1420V} or averaging over multiple exposures \citep[e.g.][]{2014PASP..126..158G, 2016A&C....16...67D} to sparsity-based signal reconstructions \citep[e.g.][]{2013A&A...557A..32R, 2017ITIP...26.5176W}.  A key advantage of hierarchical Bayesian models is the natural handling of missing data. \footnote{Also applicable to  data-sparse regimes due to shrinkage effects \citep[e.g.][]{NegBin2015}.} 

Another important missing data scenario emerges when obtaining spatially resolved data as high-resolution spectra or kinematic measurements for resolved objects (e.g. local group galaxies, molecular clouds and star forming regions, Solar System planets, etc.) is telescope-time expensive. In such situations only a few parts of the object are usually observed and analysed, and thus some kind of spatial interpolation of the derived properties is performed \citep[e.g.][]{2016ApJ...832L..23V}. In another scenario it may be necessary for an astronomer or near-future semi-autonomous instruments or surveys to take on-the-fly decisions during the night about what region of a given object needs more observations, or even if a given observation should continue based on incomplete and noisy information.

\begin{figure*}
\includegraphics[trim=0 0.1cm 0 0.1cm,clip, width=0.95\linewidth]{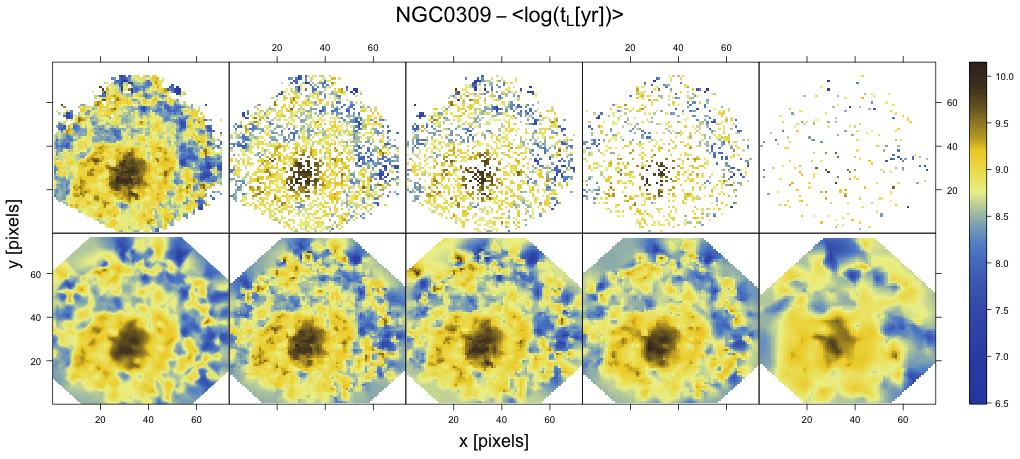}
\caption{Predictions from INLA for input {\sc starlight} age of NGC\,0309 when 100, 75, 50, 25 and 5\% of the data is used. Upper panels shows the {\sc starlight} input, bottom the INLA prediction.}
\label{sparsity1}
\end{figure*}

\begin{figure*}
\includegraphics[trim=0 0.1cm 0 0.1cm,clip,width=0.95\linewidth]{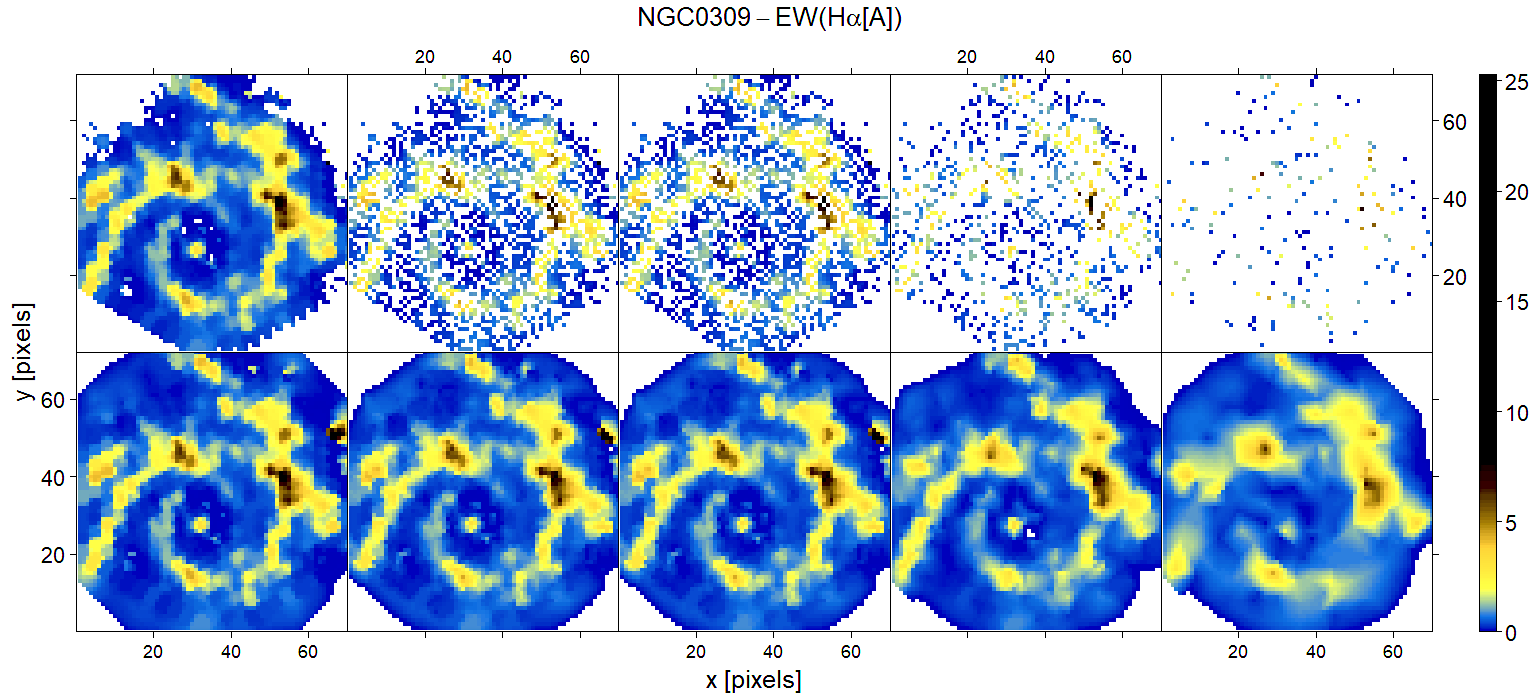}
\caption{Same as Fig. \ref{sparsity1}, but for H$\alpha$ EW retrieved from Gaussian line profile fits.}
\label{sparsity2}
\end{figure*}

\begin{figure}
\includegraphics[trim=0 0cm 0 0cm,clip,width=0.95\columnwidth]{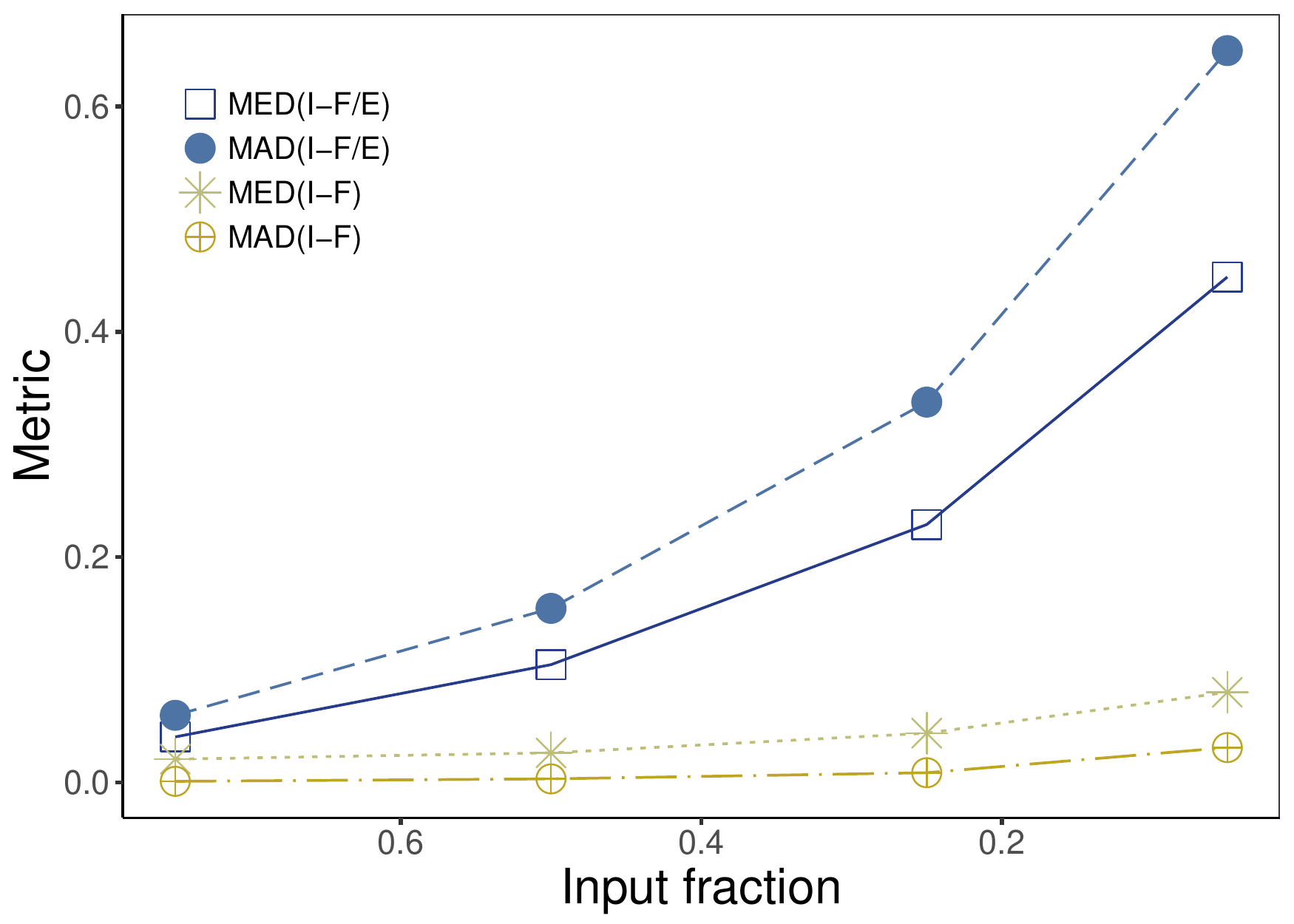}
\caption{Comparison of INLA predictions between full and fractional input age maps of NGC\,0309 for different metrics: a) the median of the squared differences between full and fractional prediction divided by the variance estimated by INLA (open square), b) the median absolute deviation (MAD) of the squared differences divided by the error (solid circle), c) the median of the raw squared difference \emph{without dividing by the error} (asterisk) and d) the MAD of the raw squared difference \emph{without dividing by the error (crossed circle)}. }
\label{metric-sparsity}
\end{figure}

Here we exemplify a result of the application of INLA to inpainting missing data, up to very sparse data regimes, by considering spatial correlations. To emulate this scenario we first randomly remove 25\%, 50\%, 75\% and 95\% of the information from the {\sc starlight} analysis of the individual IFU spaxels. Afterwards we adopt INLA to infer the underlying field, in the worst-case scenario, preserving only 5\% of the original property map pixels. Figures~\ref{sparsity1} and \ref{sparsity2} show the input data and results obtained from the field reconstructions for the age and EW(H$\alpha$) fields of NGC\,0309.
The figures clearly show that since the physical processes are spatially correlated in the observed galaxies, the INLA model is capable of reconstructing most structures of the underlying fields even when the vast majority of the data is missing. It is remarkable that in the case of both properties the majority of the large scale structures, and also a considerable fraction of the small scale structures, were reconstructed even in the worst-case scenarios considered. Figure~\ref{metric-sparsity} indicates that INLA reconstructions becomes less accurate (with respect to the reconstruction using 100\% of the data) with more missing data, as expected. However, the associated estimated uncertainties also increase, therefore the method provides conservative predictions. 

\begin{figure*}
\includegraphics[trim=2cm 0.1cm 2cm 0.1cm,clip,width=0.95\linewidth]{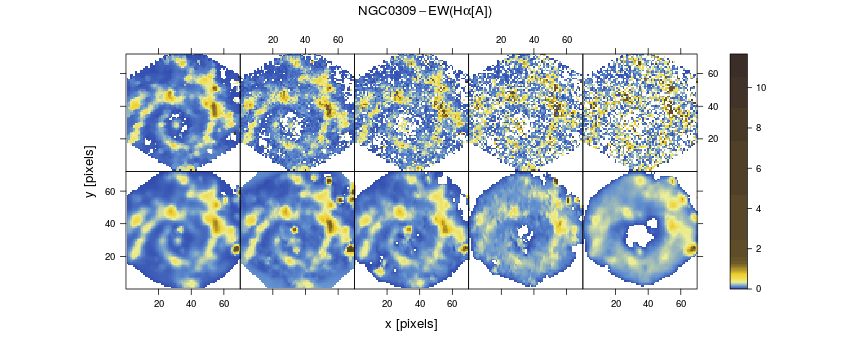}
\caption{Predictions from INLA for H$\alpha$ EW map of NGC\,0309 with S/N of 10, 2, 1, 0.5 and 0.3 (left to right). Upper panels show input and bottom panels INLA predictions. Note that the white pixels in the input data represent cases where the perturbed values in the simulation go below zero (i.e. these are points where the fitting procedure would not have converged).}
\label{signal}
\end{figure*}

We also perform a test of the INLA reconstruction at low signal-to-noise data. For this, we disturb the original H$\alpha$ map of NGC\,0309 according to a decreasing S/N ratio of 10, 2, 1, 0.5 and 0.3. Figure~\ref{signal} shows the input maps and the corresponding predicted INLA maps. These are reasonably consistent with the original high-signal map, recovering many features even at very low S/N regimes,  and without the need for spatial binning.

\subsection{Multiple pointings}

In many cases the astronomical object of interest is larger than the instrument field-of-view. This is the case of nearby extragalactic objects \citep[e.g.][]{2017ApJ...846...14B}, galaxy clusters \citep[e.g.][]{2015ApJ...804...70G}, galaxy mergers \citep[e.g.][]{2017A&A...606A..83C} and of the wide-field IFU surveys \citep[e.g.][]{2017A&A...606A..12H, 2017A&A...608A...1B}. More locally, multiple-pointing and individual observations are also used to study the interstellar medium \citep[e.g.][]{2017A&A...607A.133W}, molecular clouds and star-forming regions \citep[e.g.][]{2016MNRAS.455.4057M, 2017sfcc.confE..25G, 2018A&A...614A..65M} and correlated stellar structures within the Milky-Way, as arms and other large structures like stellar streams encompassing several degrees. At the limit case, if the data is spatially correlated in the object under study it would be possible to adopt multiple pointings of a normal spectrograph, and not an IFU, to acquire non-spatially resolved spectral data and afterwards adopt INLA to infer the spatially resolved map -- this is in fact an extreme state of missing data.

\begin{figure}
\centering
\includegraphics[trim=0.1cm 0.1cm 0.1cm 0.1cm,clip,width=0.975\linewidth]{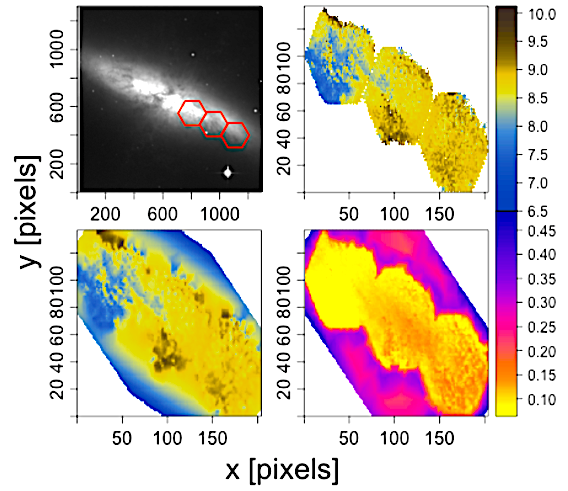}
\caption{Image of M82 in $R$-band \citep{Dale+09} with three PMAS pointings superimposed in red (\emph{top left}) from which {\sc starlight} age were obtained (\emph{top right}), the respective INLA reconstruction (\emph{bottom left}) and the INLA standard error of the posterior mean estimate  (\emph{bottom right}).}
\label{M82}
\end{figure}

If the data is spatially correlated, the reconstruction of the properties maps for such objects observed using multiple pointings of an instrument can gain significantly from the use of the INLA method. Figure \ref{M82} shows as an example the age map of the M82 galaxy. The observations were obtained from three independent PMAS pointings due to the galaxy's large angular size. The INLA reconstructed mean field shows a morphologically coherent estimation of the age map across the regions of data overlap, and that also even at extrapolated regions between pointings (the upper corners of the hexagonal FoVs). Naturally, the field of estimated uncertainties from the INLA posteriors also shows that the inference suffers at regions that are distant from the data, since the inference process lacks constraints. Moreover, this reconstruction exemplifies that INLA can provide a methodology for situations where only piecewise observation/data is available. In this particular case, instead of performing independent stellar population analysis for each pointing, it is possible to join the information into a single spatially resolved study of the object.

\section{Discussion and Conclusions}
\label{sec:conc}

Making reliable inferences from  IFU data is paramount for better exploitation of the spatial information encoded on these datasets. Regardless if these correlations are physical or caused by instrumental effects. 

Hierarchical Bayesian models provide the means to  model these spatial  properties conveying information about the probability distribution of the variable under study combined with their spatial structure.   An immediate advantage is the possibility to predict under-sampled  pixel locations, thereby naturally accommodating  missing data, bad pixels and irregular sampling. 

This work introduces the potential of spatial models via INLA for astronomers through the analysis of IFU cube data.
The  methodology assumes the galaxy properties as realizations of a random field. The model is implemented into  a spatial Bayesian framework, in which INLA is used for fast inference.\footnote{To the authors best knowledge this is the first utilization of INLA in astronomy.}

We analysed the entire CALIFA and PISCO samples by first running  a pixel-by-pixel {\sc starlight} fit,  from which we obtained the following parameters: luminosity- and mass-weighted age and metallicity, stellar mass and extinction.

We focused on two example galaxies: one elliptical and one spiral. We show that by taking into account the spatial correlations of the data in the reconstructions of spatial maps of galaxy physical properties, it is possible to recover structures in the output maps that are hidden or barely distinguishable in simple renderings of the individually fitted properties. 

We provided a comparison of our approach against existing techniques that treat spatial information in IFU cubes, namely segmentation techniques like Voronoi binning used in PyCASSO and the multi-dimensional cell partition of {\sc BaTMAn}, for the same galaxy. Although different in nature, we find clear advantages of using spatial modelling via INLA if spatial correlation exists in the data, which is expected to be the case for any spatially resolved astronomical object.

The model is capable of dealing with highly sparse regimes.  We challenged the method to analyse data-sparse regimes where up to 95\% of the information was missing. The method was capable of reconstructing most of the underlying large scale spatial structure of the galaxy properties.


Although the main focus of this work was to provide a smooth representation of galaxy properties over an irregular and noisy grid, the potential of spatial models are much wider. 
More complex structures can be seamlessly  adapted including: spatio-temporal dependencies,  multivariate maps, non-Gaussian properties\citep[e.g.][]{nel72,Bin2015,Gamma2015,2016MNRAS.461.2115D}, and so forth. Examples of potential follow-ups are: probing  the prevalence of active black holes in terms of local environment; the connection of supernova properties and their host galaxies and surroundings; to test spatial systematics in Gaia data; or for fast generation of mock datasets of spatially resolved galaxies for specific instrumental/observing conditions and redshifts of interest.

This work provides a glimpse of broad field of Bayesian spatial modelling, and  takes a step forward in the statistical analysis of astronomical IFU data.  It provides a contemporary approach for field reconstruction of galaxy properties, and a test bed for other applications (interferometry, cosmic microwave background, $N$-body simulations),  in which spatial information cannot be neglected.  

\section*{Acknowledgements}
We thank the referee for the thorough comments.
This work is a product of the $\rm 4^{th}$ COIN Residence Program (CRP\#4). We thank Emmanuel Gangler for encouraging the accomplishment of this event. CRP\#4 was held in Clermont Ferrand, France on August 2017, with support from the  Universit\'e Clermont Auvergne and the R\'egion Auvergne-Rh\^one-Alpes. 
SGG and AKM acknowledge the support from the Portuguese Strategic Programme UID/FIS/00099/2013 for CENTRA and the FCT projects PTDC/FIS-AST/31546/2017 .
RSS acknowledges the support from NASA under the Astrophysics Theory 
Program Grant 14-ATP14-0007. 
AKM acknowledges the support from the Portuguese Funda\c c\~ao para a Ci\^encia e a Tecnologia (FCT) through the grant SFRH/BPD/74697/2010 and the ESA contract AO/1-7836/14/NL/HB.
E.E.O.I acknowledges support from CNRS as part of its MOMENTUM programme over the 2018 – 2020 period.
L.G. is supported in part by the US National Science Foundation under Grant AST-1311862.
The {\sc starlight} project is supported by the Brazilian agencies CNPq, CAPES and FAPESP and by the France-Brazil CAPES/Cofecub program.
This work has made use of the computing facilities of the Laboratory of Astroinformatics (IAG/USP, NAT/Unicsul), whose purchase was made possible by the Brazilian agency FAPESP (grant 2009/54006-4) and the INCT-A. This project has been supported by a Marie Sk\l{}odowska-Curie Innovative Training Network Fellowship of the European Commission's Horizon 2020 Programme under contract number 675440 AMVA4NewPhysics. 
The Cosmostatistics Initiative\footnote{\url{https://cosmostatistics-initiative.org}}(COIN) is a non-profit organization whose aim is to nourish the synergy between astrophysics, cosmology, statistics, and machine learning communities.
COIN thanks the support from \texttt{Overleaf}\footnote{\url{https://www.overleaf.com}} collaborative platform.





\bibliographystyle{mnras}
\bibliography{biblio}



\appendix

\section{Laplace approximation}
\label{app:Laplace}
For a given probability distribution function (PDF), that is smooth around its mode $(\hat{x})$, a Laplace approximation is used to represent  the PDF with a normal distribution. It uses the 2-term Taylor series expansion around the mode of the log-pdf. If $(\hat{x})$ denotes the point of maxima of a given PDF, $p(x)$, then it is also the point of maxima of the log-pdf $ q(x) = \log{p(x)}$ and we can write: 

\begin{align}
&q(x) \approx q(\hat{x}) + (x-\hat{x})\dot{q}(\hat{x}) + \frac{1}{2}(x-\hat{x})^2\ddot{q}(\hat{x})\\
&= q(\hat{x}) +  \frac{1}{2}(x-\hat{x})^2\ddot{q}(\hat{x}),\\
&\log{p(x)} \approx cte -  \frac{1}{2}(x-\hat{x})^2\ddot{q}(\hat{x}),\\
&p(x) \approx cte\exp\left(-\frac{(x-\hat{x})^2}{2\sigma^2}\right), \quad \sigma^2=1/\ddot{q}(\hat{x})
\end{align}
Thus the Laplace approximation supposes that a given distribution is approximately Normal with mean $\hat{x}$ and variance $\sigma^2=1/\ddot{q}(\hat{x})$.


\bsp	
\label{lastpage}
\end{document}